# Round table 2 – Clinical research and methodology

# What usage and what hierarchical order for secondary endpoints?

Silvy Laporte[1], Marine Diviné[2], Danièle Girault[3] and the participants at the Giens round table no. 2: Pierre Boutouyrie [4], Olivier Chassany [5], Michel Cucherat [6], Herve de Trogoff [7], Sophie Dubois [8], Cecile Fouret [9], Natalie Hoog-Labouret [10], Pascale Jolliet [11], Patrick Mismetti [1], Raphaël Porcher [12], Cécile Rey-Coquais [13], Eric Vicaut [14].

1. Jean-Monnet University, Inserm UMR1059, Clinical Research Unit, Innovation, Pharmacology, Saint-Etienne University Hospital, France
2. Amgen, Boulogne Billancourt, France
3. Denelia, Clichy, France
4. AP-HP (Paris Public Hospital System), France
5. Paris Diderot University, EA 7334 REMES, Clinical Research Unit URC-ECO, AP-HP, France
6. UMR CNRS 5558, Lyon 1 University, Lyon, France
7. Astellas Pharma, Levallois-Perret, France
8. Takeda, Puteaux, France
9. Medtronic, Boulogne Billancourt, France
10. Innovation and Research Centre, INCA, Boulogne-Billancourt, France
11. Faculty of Medicine, University of Nantes, France
12. The Centre of Research in Epidemiology and Statistics Sorbonne Paris Cité (CRESS – Inserm UMR1153), AP-HP, France
13. Pfizer, Paris, France
14. Saint-Louis Lariboisière Hospital, AP-HP, France

**Key words**

Endpoint; secondary endpoints; hierarchical test procedure; inflation of type I error

**Summary (140 words)**

In a randomised clinical trial, when the result of the primary endpoint shows a significant benefit, the secondary endpoints are scrutinised to identify additional effects of the treatment. However, this approach entails a risk of concluding that there is a benefit for one of these endpoints when such benefit does not exist (inflation of type I error risk). There are mainly two methods used to control the risk of drawing erroneous conclusions for secondary endpoints. The first method consists of distributing the risk over several co-primary endpoints, so as to maintain an overall risk of 5%. The second is the hierarchical test procedure, which consists of first establishing a hierarchy of the endpoints, then evaluating each endpoint in succession according to this hierarchy while the endpoints continue to show statistical significance. This simple method makes it possible to show the additional advantages of treatments and to identify the factors that differentiate them.

# 1. Introduction

Imagine two medicinal products assessed in the treatment of non-small cell lung cancer in two randomised clinical trials. Both molecules show an improvement in progression-free survival – the primary endpoint – compared to the same comparator. In one of the two studies, a difference in overall mortality is also identified. The question then arises of how to illustrate this differential result. Similarly, consider a direct oral anticoagulant assessed in the treatment of pulmonary embolism. Given its convenience compared to the comparator, the objective of the study was to demonstrate the non-inferiority of this anticoagulant compared to a vitamin K antagonist for a switch from low molecular weight heparin. The study not only made it possible to demonstrate non-inferiority compared to the primary endpoint (recurrence of clinical events under treatment), but also evidenced a significant reduction in the risk of major bleeding with the new anticoagulant [1]. Can this result be used, and the anticoagulant considered therapeutic progress in terms of treatment safety?

From the clinician's or manufacturer's point of view, differentiating the result highlighted on the secondary endpoint is important and seems valid since the primary endpoint is significant. It can therefore be used to select the optimal treatment for patients or to claim a higher improvement of actual benefit (ASMR/IAB) than existing therapies. However, from a methodological standpoint, due to the numerous endpoints, drawing conclusions based on the secondary endpoints induces a risk of erroneously concluding that there is an additional treatment benefit when in fact there is not. In order to reconcile the two viewpoints, we will discuss the regulatory and scientific admissibility of the results for the secondary endpoints and propose analytical methods for these endpoints in view of registration, assessment (Health Technology Assessment), communication and recommendations.

## 2. The issue with secondary endpoints: there is a risk of erroneously identifying a benefit

In the process of assessing a new treatment, the results of therapeutic trials lead to different types of conclusions regarding the benefit(s) of the treatment. Generally speaking, the primary objective is to demonstrate a potential clinical benefit of the treatment, which could lead to the treatment being considered for use in medical practice. However, once the existence of a treatment benefit has been concluded, the question arises of whether the treatment has other effects that may increase its clinical interest. For example, if the treatment assessment is based on a reduction of morbidity and mortality, a conclusion of an additional reduction in mortality maximises the treatment's clinical interest and may lead to it becoming the reference treatment in its class. Given the existence of random sampling fluctuations, there is a risk of erroneously drawing these conclusions solely by chance and therefore, in statistical terms, of committing a type I error. To control the type I error risk, conclusions are drawn based on a statistical test, conventionally conducted with a bilateral 5% significance threshold. The risk of drawing an erroneous conclusion is therefore less than or equal to 5%.

However, if the trial shows a treatment effect on several endpoints simultaneously, type I error risk will be inflated. Though the initial conclusion that at least one treatment benefit exists may be drawn from the simultaneous evaluation of the $p$ values for several endpoints, the risk of erroneously concluding that a treatment benefit exists is no longer 5%, but much greater. Evaluation of the 1$^{st}$ endpoint may result in wrongly concluding the existence of a treatment benefit with a 5% risk, to which is added, if a conclusion cannot be drawn from this endpoint, a 5% risk of error when evaluating the 2$^{nd}$ endpoint, and so on. Ultimately, even under the (null) hypothesis that the treatment has no effect on any of the endpoints, if enough tests are conducted, it is possible to find one that makes it possible to erroneously conclude that there is a treatment benefit. To avoid this inflation and completely control the risk of type I error when assessing the potential benefit of a treatment, the conventional solution is to make the decision based on one single test, by predefining a primary endpoint, which should be the sole basis of the decision-making process. After analysis of the trial, it would therefore be impossible to select false positive results obtained by chance as proof of a treatment benefit.

In practice, this approach does not prevent the search for an additional benefit when reviewing secondary endpoints. The multiplicity of the secondary endpoints induces an increased risk of type I error when assessing additional benefit. With enough (independent) secondary endpoints, finding an additional advantage of the treatment becomes very likely, even if in reality there are none. For example, in the absence of type I error control, it is possible, in the presence of multiple drugs with the same mechanism of action, to erroneously draw conclusions as to differences in clinical benefit and select a reference treatment based on incorrect assumptions.

## 3. One single endpoint for one benefit?

In order to avoid the problems raised by the multiplicity of the tests, textbooks [2,3] and methodology guides from regulatory agencies [4] recommend that a randomised clinical trial have a unique primary endpoint. In this case, the efficacy endpoint for the investigational treatment is based entirely on the analysis of the primary endpoint. The other efficacy endpoints collected and analysed – called secondary endpoints – do not allow for official recognition of an additional treatment benefit. In order to avoid inflation of type I error risk, the primary endpoint must evidently be predetermined in the protocol, and the presentation of the study results must comply with the protocol. Several studies have shown differences between the primary endpoints reported in the publications and those set out in the protocol [5] or indicated on clinical trials registries such as www.clinicatrials.gov [6,7].

Although this paradigm allows for the simple control of type I error risk and facilitated interpretation of trial results, it has limitations. In particular, it supposes that a single variable is enough to demonstrate a pertinent treatment effect. However, it is often too simplistic to summarise the potential benefit of a treatment in a single endpoint. For example, in the treatment of chronic obstructive pulmonary disease (COPD), it is recommended that efficacy trials include forced expiratory volume in 1 second (FEV1) both as a measure of pulmonary function and of symptomatic benefit [8–9]. In cancerology, for advanced disease, it is also the norm to use overall survival and progression-free survival as endpoints.

In these cases, a set of primary or co-primary endpoints can be defined. One way to avoid inflation of the risk of a type I error is to specify that a treatment benefit will only be said to exist if the test of each of the co-primary endpoints is significant at the predefined significance threshold. This is the approach recommended by the EMA's Committee for Proprietary Medicinal Products (CPMP) in the above-mentioned example of COPD. In this case, no reduction of the $\alpha$ threshold is required to control the risk of overall type I error. However, this procedure increases the risk of overall type II error in the trial (that is, the probability of overlooking a benefit that does exist), and should therefore lead to increasing the size of the trial to reach the desired power.

Requiring that several primary endpoints be significant can be justified if one considers that one of these endpoints is not enough to declare a medicinal product effective. In other cases, however, one of these co-primary endpoints is sufficient. For example, a cancer treatment can be considered as having a benefit if it increases overall or progression-free survival (though this position is debatable, it is seen in practice [10,11]). One possible solution is thus to distribute the risk of type I error between the different co-endpoints. In this type of trial, superiority can be concluded based on one of the endpoints, even when no benefit is demonstrated in regard to the other co-primary endpoints, in contrast to the previous case. Conventional approaches for multiple tests (e.g. Holm or Hochberg) may then be used. The simplest method is distribution between the different co-primary endpoints so the cumulative risk induced with each endpoint does not exceed the overall type I error when assessing treatment benefit. For example, in a trial comparing bevacizumab to placebo in the treatment of glioblastomas, overall survival was tested with a threshold of 0.046 and progression-free survival with a threshold of 0.004 [12]. In other cases, an equal division between the co-primary endpoints – 0.025 for each if there are two – can be considered. This threshold adjustment for the tests also involves an increase in the trial size to preserve its power, especially as the number of endpoints tested increases. For example, to test five co-primary endpoints with an equal risk distribution, tests should be conducted with a threshold of 0.01 and about 20% more subjects than would be used in a situation considering only two co-primary endpoints. Other corrective methods, sometimes substantially more complex, have been proposed [13–15]. A simpler method, the hierarchical test procedure, may also be used.

## 4. The hierarchical test procedure

The principle behind this method is simple (Figure 1) [16,17]. Endpoints are predefined in the protocol in a definitive hierarchical manner (first, second, etc.). The first endpoint is analysed according to the method for testing a single primary endpoint. The risk of erroneously identifying a treatment benefit is completely controlled (limited to 5% for simplicity). If the test result is not significant, the procedure is stopped and the treatment is not considered useful based on this trial. If the test result is significant, the question arises of whether there is at least one additional benefit. Since the endpoint to be used to search for additional benefits is pre-established in the hierarchy, no 'fishing for results' occurs. The hypothesis that the treatment has a first additional benefit is tested with a completely controlled 5% risk. If this test shows statistical significance, it can be concluded that the treatment has a benefit ($1^{st}$ test) as well as an additional benefit ($2^{nd}$ test). If a treatment has shown both a primary benefit and first additional benefit of the treatment, the question of whether there is a second additional benefit is raised. To avoid 'fishing for results' from all the remaining secondary endpoints, the hierarchical test procedure perfectly defines the single endpoint, which should be evaluated to search for this second additional benefit, therefore preventing any inflation of the type I error. This process continues until the end of the hierarchy or until the first non-significant test. No conclusions can be drawn before the $1^{st}$ non-conclusive test ($p < 0.05$), even with a $p$ value $< 0.05$, because it would constitute a fishing approach involving the same traps as those observed during analyses of secondary endpoints.

The benefit of this approach is to potentially be able to demonstrate an effect on several endpoints from a single trial based on the results obtained, without inflation of the type I error. Due to the interdependent nature of the tests, the hierarchical test procedure allows for a perfect control of the risk of false discovery when assessing potential further treatment benefits. This method has thus been widely used since its 'consecration' in 2002 by the recommendations of the European registration authority on the multiple statistical tests [18].

Lastly, the method can be combined with other adjustments (cf. PLATO frame), and can involve composite endpoints, subgroups, etc. For example, the 1st hierarchy test may seek to prove non-inferiority on a certain endpoint, the 2nd to prove superiority on another endpoint, the 3rd to demonstrate the superiority of this endpoint specifically in a subgroup of high interest patients (because a competitor has specifically shown interest in this population), etc.

The hierarchical test procedure is the only method currently available that can demonstrate additional benefits of treatments. It provides the possibility of obtaining several elements of statistical evidence from one trial.

**Focus: the PLATO study [15,19]**

PLATO was the first major study in which this method was applied. This randomised clinical trial compared two anti-platelet strategies combined with aspirin (ticagrelor *versus* clopidogrel) in the management of 18,624 patients with acute coronary syndrome (ACS).

The statistical analysis plan for this study focused first on a double analysis of the primary endpoint in two study populations (the total population and the subpopulation with angioplasty), with the statistical significance being retained if and only if each analysis came out simultaneously at the 5% threshold to avoid any inflation of the type I error (Table). The choice of these two analyses allowed the manufacturer to claim an indication in ACS, regardless of its management, and more specifically in the ACS subpopulation managed invasively with angioplasty. Once this efficacy endpoint was analysed for the two different populations, a hierarchical analysis was planned so as to claim an additional benefit for several other efficacy endpoints, with the analysis following a predetermined order (Table).

| Endpoints | Order of analysis | ticagrelor N = 9333 | clopidogrel N = 9291 | p |
|---|---|---|---|---|
| **Primary endpoint**: cardiovascular death + MI + stroke, total population | 1 | 9.8 % | 11.7 % | <0.001 |
| **Primary endpoint**, scheduled primary angioplasty population | 1 | 8.9 % | 10.6 % | 0.003 |
| All-cause mortality + MI + stroke | 2 | 10.2 % | 12.3 % | <0.001 |
| Cardiovascular death + MI + stroke + refractory ischaemia + other arterial events | 3 | 14.6 % | 16.7 % | <0.001 |
| MI | | | | |
| Cardiovascular death | 4 | 5.8 % | 6.9 % | 0.005 |
| | 5 | 4.0 % | 5.1 % | 0.001 |
| Stroke | 6 | 1.5 % | 1.3 % | **0.22 STOP** |
| All-cause mortality | 7 | 4.5 % | 5.9 % | <0.001 |
| CP from D0 to D30 | 8 | 4.8 % | 5.4 % | 0.045 |
| CP from D31 to D360 | 9 | 5.3 % | 6.6 % | <0.001 |
| Stent thrombosis | 10 | | | <0.01 |

According to the hierarchical test procedure, the statistical tests were to be performed as long as the endpoints, taken in order, were significant at the 5% threshold. Thus, in the PLATO study, the statistical tests should have theoretically stopped upon analysis of the 6$^{th}$ endpoint (strokes, $p = 0.22$), and the following assessment criteria in the hierarchy should only be presented in a descriptive way, as in the case of a conventional analysis of secondary endpoints. However, the statistical tests were continued beyond the 1$^{st}$ non-significant endpoint and the results were even highlighted in the publication, particularly as regards total mortality, even when the risk of erroneous conclusion was no longer controlled. The publication of the PLATO study results illustrates the need to raise awareness regarding the method and the interpretation of results, and fully justifies this issue having been the subject of a round table.

## 5) How to construct the hierarchy

During the analysis of the trial, hypotheses are tested one after the other, following a pre-specified hierarchy defined in the protocol. Not all endpoints with $p < 0.05$ are considered significant: an effect will be considered to have been demonstrated for all endpoints obtaining a $p < 0.05$ (5%) until the first endpoint for which $p > 0.05$ in the order of the hierarchy. The order of the endpoints indicates a strategic choice.

## 5.1 How to select endpoints to include in the hierarchy

The choice of endpoints can remain relatively simple insofar as the hierarchy is focused on the endpoints for which a benefit is sought. As with any choice regarding assessment criteria, the endpoints have to be selected in relation to the question posed and thus to the relevant therapeutic objective for the patient. Safety endpoints may also be considered in this hierarchy when an additional benefit is expected in terms of risk. Examples are provided by the European [9] and American [20] guidelines (cf. frame). The challenge is then to determine the order of the hierarchy in the endpoint model. Lastly, from a statistical point of view, once they are ordered hierarchically, all endpoints selected have the same level in terms of statistical evidence.

> **Example provided by the European guidelines for chronic obstructive pulmonary disease (excerpt from [9]).**
> The hierarchical order is given as an example. The choice of endpoints might be influenced by the mechanism of action of the test product.
> 1. Lung function, i.e. FEV1
> 2. Number of exacerbations
> 3. Dyspnoea (Breathlessness) using for example the TDI (Transition Dyspnoea Index)
> 4. Disease-specific Health-Related Quality of Life questionnaire (e.g. SGRQ)
> 5. Assessment of exercise capacity
> 6. Rescue medication
> 7. Other symptoms (e.g. chest symptoms, cough/sputum)
> 8. Other lung measures*
>
> \* Lung function (other than $FEV_1$) rarely showed improvement with current symptomatic COPD treatments. It is thus logical to place it at the end of the hierarchical endpoint model.
>
> Lung function can be the primary endpoint, "**but if alone, is considered insufficient in the assessment of therapeutic effect**". So there is a need for efficacy demonstration on other important endpoints.
>
> It can be also a co-primary endpoint with lung function and number of exacerbations or dyspnoea or disease specific questionnaire or exercise capacity. "**Efficacy should be demonstrated convincingly for both co-primary endpoints and improvements seen in these endpoints must be statistically significant and clinically relevant**."

## 5.2 How to order the endpoints in the hierarchy

The choice for the 1st endpoint in the hierarchy follows the same rules as those regarding the choice for the primary endpoint: carefully selected to correspond to a relevant clinical endpoint with regard to the therapeutic objective for the disease, but sufficiently common to enable the study to be conclusive on this endpoint. The calculation of the required number of subjects is based on the frequency of the event and the magnitude of the expected effect for this endpoint. For the rest of the hierarchy, frequent events can be preferred as long as these selected endpoints are considered to be of clinical interest. It is recommended that the number of subjects required for each endpoint of the hierarchy be calculated in order to assess the size of the study in relation to the expected results, or to redefine the hierarchy. The potential lack of power for the assessment of the endpoints of the hierarchy is to be taken into account when constructing the hierarchy. A compromise must therefore be found to order these pertinent endpoints while staying focused on the benefit to the patients.

For example, a composite endpoint for morbidity and mortality is easier to select than total mortality as its base frequency is higher. Therefore, events that are less frequent (but more clinically relevant) should be considered further down the hierarchy, given the decreased power and accuracy of the estimated effect.

Ultimately, the order of the hierarchy is not overly important, the objective being to obtain results that demonstrate the clinical benefit of a treatment in terms of efficacy and possibly safety.

**5.3 Discuss the hierarchy with the Healthcare Authorities**

In the case of studies on medicinal products, the European Medicines Agency (EMA) can be contacted for Scientific Advice [21] regarding the choice and order of the endpoints. The Scientific Advice procedure was put in place by the EMA to advise manufacturers on the developmental or experimental aspects of clinical trials. Its advisory opinions are without prejudicing the outcome of obtaining a Marketing Authorisation. Scientific Advice (Shaping European Early Dialogues) can also be requested with the HTA (Health Technology Assessment) agencies, or conjointly with the EMA and the HTA agencies (parallel EMA multi HTA early dialogue).

> **Focus: the PLATO study (cont'd)**
>
> The PLATO study is all the more exemplary as it illustrates the importance of the choice of hierarchical sequence. Retrospectively (which is always easier!), the choice of putting stroke at the start of the hierarchy was risky, as this endpoint combines both ischaemic stroke (efficacy) and haemorrhagic stroke (safety in use). However, as is often the case with antithrombotic treatments, an expected gain in terms of efficacy comes at the price of increased haemorrhagic risk. Therefore, the chances of obtaining a significant test result for an endpoint combining efficacy and safety are virtually nil. Moreover, it is worth noting that the incidence of this endpoint is lower than that of all other endpoints, and the power of the analysis was, from the outset, more limited than for the following endpoints.

## 6. Recommendations

The round table has issued the following recommendations so that additional benefits relating to the secondary endpoints can be claimed and recognised.

6.1. Use the hierarchical test procedure in randomised interventional studies so as to be able to claim the additional benefits demonstrated for the secondary endpoints.

6.2. If an already-initiated study would like to implement a hierarchical test sequence that was not initially scheduled, an amendment to the protocol may be considered provided that the decision is made before unblinding of the data.

6.3. Respect the methodological rules regarding statistical tests: claimed results for endpoints in the hierarchy are legitimate as long as the endpoints remain significant in the order of the hierarchy. After the $1^{st}$ non-significant test, results can only be descriptive.

6.4. The hierarchical test procedure is entirely admissible at a statistical level with controlled monitoring of the risk of overall error; however, the use of this method does not mean that discussion of possible biases in the study, the clinical relevance of the selected endpoints, or the significance of the effects observed for the selected endpoints can be overlooked.

6.5. Not all secondary endpoints need to be included in the hierarchy. Secondary endpoints not included in the hierarchy can be considered. There are aspects of clinical trials for which claims cannot be made. The results for these endpoints will therefore be descriptive only.

6.6. Training for the method is crucial in order to obtain a legitimate interpretation of the results of the approach. Particular care is recommended for the critical reading of the article and the study results (cf. PLATO frame), both during the presentation of the study results to the authorities and during their use by the manufacturer in a promotional context.

## 7. Conclusions and perspectives

The issue of secondary endpoints in clinical trials may arise generally in two different forms. In the first, the description of the treatment effect on the pathology requires several parameters, which leads to the consideration of several criteria for the primary endpoint of the study. We have therefore shown that the use of co-endpoints is possible if they are accompanied by the essential type I error adjustments. In the second, the secondary endpoints make it possible to differentiate treatments that demonstrate an effect on the primary endpoint but that have different levels of proof on additional effects. In this case, the hierarchical test procedure is the only simple method currently available that can demonstrate a medical benefit on several endpoints.

**Figure 1. Method outline**

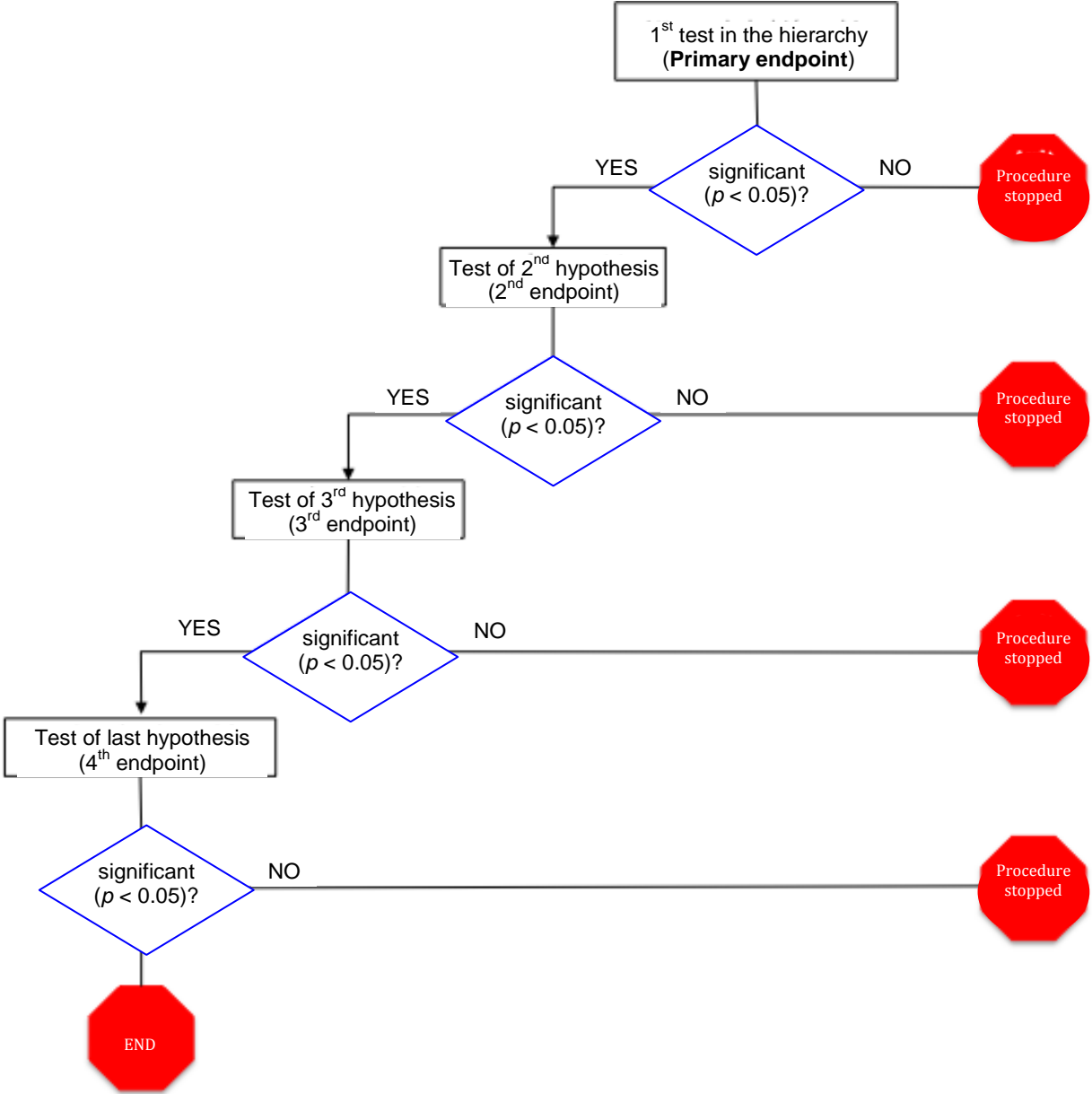

## List of abbreviations

| ACS | Acute Coronary Syndrome |
| --- | --- |
| ASMR/IAB | Improvement in actual benefit |
| COPD | Chronic obstructive pulmonary disease |
| CPMP | Committee for Proprietary Medicinal Products |
| EMA | European Medicine Agency |
| FDA | Food and Drug Administration |
| FEV1 | Forced expiratory volume in 1 second |
| HTA | Health Technology Assessment |

## Conflicts of interests

None

## Correspondence and offprints

Prof. Silvy Laporte, Jean-Monnet University, Inserm UMR1059, Clinical Research Unit, Innovation, Pharmacology, CHU Saint-Etienne, France